\documentclass[12pt]{raa}            
\usepackage{graphicx,times}
\usepackage{natbib}

\providecommand{\bm}[1]{\mbox{\boldmath$#1$\unboldmath}}

\usepackage{epsfig}
\usepackage{txfonts}
\usepackage{lscape}
\begin{document}

\title{Central gas entropy excess as a direct evidence for AGN feedback in galaxy groups and clusters}

\author{Yu Wang\inst{1}, 
Haiguang Xu\inst{1},
Junhua Gu\inst{1},
Liyi Gu\inst{1},
Jingying Wang\inst{1}
and
Zhongli Zhang\inst{2}
}

\institute{
$^{1}$Department of Physics, Shanghai Jiao Tong University,
800 Dongchuan Road, Shanghai 200240, China; 
{\it wenyu$_{-}$wang@sjtu.edu.cn}; {\it hgxu@sjtu.edu.cn}
$^{2}$Max-Planck-Institute for Astrophysics, Karl-Schwarzschild-Str. 1, 
Postfach 1317 D-85741, Garching, Germany}

\abstract{
By analyzing {\it Chandra} X-ray data of a sample of 21 galaxy groups 
and 19 galaxy clusters, we find that in 31 sample systems there exists 
a significant central ($R^{<}_{\sim} 10h_{71}^{-1}$ kpc) gas entropy excess 
($\Delta K_{0}$), which corresponds to $\simeq0.1-0.5$ keV per gas particle,
beyond the power-law model that best fits the radial entropy profile of 
outer regions. We also find a distinct correlation 
between the central entropy excess $\Delta K_{0}$ and 
$K$-band luminosity $L_{K}$ of the central dominating galaxies (CDGs), 
which is scaled as $\Delta K_{0}\propto L_{K}^{1.6\pm0.4}$,
where $L_{K}$ is tightly associated with the mass of the supermassive black hole 
hosted in the CDG. In fact, if an effective mass-to-energy 
conversion-efficiency of 0.02 is assumed for the accretion process,
the cumulative AGN feedback $E^{\rm AGN}_{\rm feedback}\simeq\eta M_{\rm BH}c^{2}$ 
yields an extra heating of $\simeq0.5-17.0$ keV per particle, 
which is sufficient to explain the central entropy excess.
In most cases the AGN contribution can compensate 
the radiative loss of the X-ray gas within the cooling radius 
($\simeq0.002-2.2$ keV per particle), and apparently exceeds 
the energy required to deviate the scaling relations from 
the self-similar predictions ($\simeq0.2-1.0$ keV per particle).
In contrast to the AGN feedback, the extra heating provided by 
supernova explosions accounts for $\simeq0.01-0.08$ keV per particle 
in groups and is almost negligible in clusters. 
Therefore, the observed correlation between $\Delta K_{0}$ and $L_{K}$ 
can be considered as a direct evidence 
for AGN feedback in galaxy groups and clusters.  
}

\authorrunning{Y. Wang et al.}  
\titlerunning{AGN Feedback in Galaxy Groups and Clusters} 
\maketitle

\section{INTRODUCTION}
Over the past few decades many observational and theoretical efforts 
have been devoted to the studies of galaxy groups and clusters. 
As of today, however, some fundamental astrophysical processes
that determine the basic properties of these celestial systems
are still poorly understood. For example, the scaling relations 
between X-ray luminosity ($L_{X}$), gas temperature ($T_{X}$), 
gas entropy ($K$), and total gravitating mass ($M_{\rm total}$), 
which are predicted by the self-similar gravitational collapse 
scenario (e.g., Kaiser 1986; Navarro, Frenk \& White 1995), are 
challenged by observed deviations in such a distinct way that 
non-gravitational heating sources are invoked to dominate the gas 
heating process in the central $\simeq$100 kpc 
(e.g., Ponman et al. 2003; Donahue et al. 2006; Sun et al. 2009). 
Currently, most of the efforts have been focused on the AGN heating 
of the inter-galactic medium (IGM) (see McNamara \& Nulsen 2007 for a review), 
since it is estimated that powerful AGN outbursts may repeat 
per $10^{6}-10^{8}$ yr and release $10^{58}-10^{62}$ ergs per outburst 
into the environment, and this amount of energy is sufficient to 
balance gas cooling and heating on group and cluster scales 
(e.g., Rafferty et al. 2006, 2008; Birzan et al. 2009). 

However, there also exist problems in the AGN feedback scenario. 
For example, the absence of X-ray cavities, a natural product of 
the AGN activity, has been reported in about 40\% cool core systems 
(e.g., Cavagnolo et al. 2008; Birzan et al. 2009), and only about 
10\% of quasars are found to host powerful radio jets 
(e.g., White et al. 2007), which indicate that cavity- and jet-related 
feedbacks might not be generic. In NGC 4051, the observed mass and energy 
outflow rates due to the AGN activity is $4-5$ orders of magnitude 
below those required for efficient feedback (Mathur et al. 2009). 
Also, Jetha et al. (2007) reported that no significant difference
of gas entropy profiles between radio-loud and radio-quiet galaxy 
groups is found (see also Sun et al. 2009).

If AGN feedback does dominate the gas heating history in 
central regions of galaxy groups and clusters, it should 
be proportional to the central gas entropy excess 
$\Delta K$, i.e., $E^{\rm AGN}_{\rm feedback}\propto \Delta K$ 
(Voit \& Donahue 2005), where $\Delta K$ is measured beyond 
a power-law model that best describes the gas entropy distribution 
in the intermediate and outer regions. On the other hand, 
by studying AGN cavities embedded in the X-ray halos of 
galaxy groups and clusters, Allen et al. (2006) found a close 
relation between the AGN feedback energy $E^{\rm AGN}_{\rm feedback}$ 
and the Bondi accretion power 
$P_{\rm Bondi}=\eta \dot{M}c^{2} \propto M_{\rm BH}^{2}$
($\eta$ is the mass-to-energy conversion efficiency, $\dot{M}$ 
is the accretion rate, and $M_{\rm BH}$ is the black hole mass; 
Bondi 1952), which indicates that the AGN feedback should 
scale with the SMBH mass in the form of 
$E^{\rm AGN}_{\rm feedback} \propto M_{\rm BH}^{2}$. 
And we know that $M_{\rm BH}$ is related to the galaxy's bulge 
luminosity $L_{\rm bulge}$ and thus galaxy's $K$-band luminosity 
$L_{K}$ via $M_{\rm BH} \propto L_{\rm bulge} \propto L_{K}$ 
(Marconi \& Hunt 2003; Batcheldor et al. 2007).
Given above relations, we expect a tight correlation between 
the central gas entropy excess and the galaxy's $K$-band 
luminosity, i.e., $\Delta K \propto L_{K}^{2}$, which has never 
been reported in literature.

In order to examine whether this correlation holds or not, 
we analyze the {\it Chandra} archive data of a sample of 
21 galaxy groups and 19 galaxy clusters to measure the central 
gas entropy excesses, and then compare them with the 
$K$-band luminosities of central dominating galaxies (CDGs). 
In \S2 we describe our sample, {\it Chandra} 
observations, and data reduction. In \S3 we measure gas density 
and temperature, and study the central gas entropy excess
against $L_{K}$. In \S4 and \S5 we discuss and summarize our results,
respectively. Throughout the paper, we adopt the cosmological parameters
$H_{0} = 71$ km s$^{-1}$ Mpc$^{-1}$,
$\Omega_{\rm b} = 0.044$,
$\Omega_{\rm M} = 0.27$,
and
$\Omega_{\Lambda} = 0.73$.
Unless stated otherwise, the quoted errors stand for the 90$\%$ 
confidence limits.

\section{SAMPLE, OBSERVATION, AND DATA REDUCTION}
\subsection{Sample selection}
In order to achieve our scientific goal, we need to select bright, 
nearby galaxy groups and clusters whose central galaxy can be well 
resolved with the {\it Chandra} Advanced CCD Imaging Spectrometer 
(ACIS). To investigate the interaction between the central AGN and 
the IGM, the selected systems are limited to have only one bright 
central-dominating galaxy and show no significant merger 
signatures. Our sample consists of 20 brightest nearby ($z<0.1$) 
galaxy groups and clusters selected from the flux-limited {\it ASCA} 
sample of Ikebe et al. (2002), and 18 galaxy groups from two 
{\it ROSAT} group samples constructed by Mulchaey et al. (2003) 
and Osmond \& Ponman (2004), respectively. Besides, we add 
the group NGC 3402 ($z=0.0153$), which satisfies the sample 
selection criteria, and the giant AGN cavity cluster MS 0735.6+7421 
($z=0.216$) for comparison. All the selected systems lie above 
the Galactic latitudes of $\pm15^{\circ}$, and are located outside 
the fields of the Magellanic Clouds. Some basic properties of 
the sample members are summarized in Table 1.

\begin{table}[h!!!]
\caption{The sample}
\begin{center}
\tiny
\begin{tabular}{lcccccccc}
\hline\hline
Group/cluster&Redshift&CDG&log($L_{B}/{\rm L_{B,\odot}}$)$^{~\it a}$&log($L_{K}/{\rm L_{K,\odot}}$)$^{~\it b}$&&
\multicolumn{3}{c}{Chandra observation}\\ 
\cline{7-9}
&&&&&&ObsID&Date& Raw (net) exposure (ks)\\
\hline
NGC 0383 &0.0173&NGC 0383 &10.86 &11.71$\pm$0.01&&2147&2000-11-06&47.2(44.4)\\ 
NGC 0507 &0.0170&NGC 0507 &11.02 &11.76$\pm$0.01&&2882&2002-01-08&43.7(43.4)\\ 
NGC 0533 &0.0181&NGC 0533 &11.07 &11.75$\pm$0.01&&2880&2002-07-28&37.6(37.4)\\  
NGC 0720 &0.0059&NGC 0720 &10.49 &11.25$\pm$0.01&&7372&2006-08-06&49.4(48.8)\\  
NGC 0741 &0.0185&NGC 0741 &11.19 &11.84$\pm$0.01&&2223&2001-01-28&30.4(29.6)\\ 
NGC 1407 &0.0057&NGC 1407 &10.77 &11.46$\pm$0.01&& 791&2000-08-16&48.6(48.6)\\
NGC 2563 &0.0159&NGC 2563 &10.59 &11.42$\pm$0.01&&7925&2007-09-18&48.8(47.9)\\
NGC 3402 &0.0153&NGC 3402 &10.76 &11.42$\pm$0.01&&3243&2002-11-05&29.5(29.3)\\  
NGC 4125 &0.0050&NGC 4125 &10.70 &11.28$\pm$0.01&&2071&2001-09-09&64.2(63.6)\\  
NGC 4261 &0.0068&NGC 4261 &10.72 &11.38$\pm$0.01&&9569&2008-02-12&101.0(99.3)\\
NGC 4325 &0.0254&NGC 4325 &10.66 &11.30$\pm$0.02&&3232&2003-02-04&30.1(29.9)\\
NGC 5044 &0.0082&NGC 5044 &10.85 &11.37$\pm$0.01&&9399&2008-03-07&82.7(82.7)\\  
NGC 5129 &0.0233&NGC 5129 &11.07 &11.65$\pm$0.01&&6944&2006-04-13&21.0(21.0)\\  
&&&&&&7325&2006-04-14&26.7(25.8)\\  
NGC 5846 &0.0063&NGC 5846 &10.70 &11.45$\pm$0.01&&7923&2007-06-12&90.4(89.8)\\  
NGC 6269 &0.0353&NGC 6269 &11.38 &11.95$\pm$0.01&&4972&2003-12-29&41.4(39.6)\\  
NGC 7619 &0.0116&NGC 7619 &10.98 &11.54$\pm$0.01&&3955&2003-09-24&37.5(28.8)\\  
HCG 42   &0.0128&NGC 3091 &10.88 &11.60$\pm$0.01&&3215&2002-03-26&31.7(31.7)\\  
MKW 3S   &0.0450&NGC 5920 &11.02 &11.57$\pm$0.02&& 900&2000-04-03&59.1(57.2)\\  
MKW 4    &0.0201&NGC 4073 &11.14 &11.83$\pm$0.01&&3234&2002-11-24&30.0(29.9)\\  
MKW 4S   &0.0286&NGC 4104 &11.20 &11.97$\pm$0.01&&6939&2006-02-16&37.1(35.8)\\  
UGC 12064&0.0166&UGC 12064&10.48 &11.12$\pm$0.01&&4057&2003-09-18&29.2(21.4)\\  
2A0335+096&0.0349&PGC 013424&11.21 &11.83$\pm$0.02&& 919&2000-09-06&21.4(19.7)\\  
Abell 0085&0.0556&PGC 002501&11.23 &12.05$\pm$0.02&& 904&2000-08-19&40.5(38.4)\\  
Abell 0262&0.0161&NGC 0708  &10.60 &11.61$\pm$0.01&&7921&2006-11-20&111.9(110.7)\\  
Abell 0478&0.0900&PGC 014685&11.32 &12.04$\pm$0.03&&1669&2001-01-27&42.4(42.4)\\  
&&&&&&6102&2004-09-13&10.4(9.9)\\
Abell 0496&0.0328&PGC 015524&11.18 &11.87$\pm$0.02&&4976&2004-07-22&75.1(63.6)\\  
Abell 0780&0.0538&3C 218    &11.38 &11.70$\pm$0.03&&4970&2004-10-22&100.7(98.8)\\  
Abell 1651&0.0850&PGC 088678&11.00 &12.42$\pm$0.04&&4185&2003-03-02&9.7(9.6)  \\  
Abell 1795&0.0616&PGC 049005&11.01 &11.93$\pm$0.03&& 493&2000-03-21&21.3(19.6)\\  
Abell 2029&0.0767&IC 1101   &11.33 &12.31$\pm$0.02&&4977&2004-01-08&77.9(77.7)\\  
Abell 2052&0.0348&3C 317    &10.87 &11.87$\pm$0.02&&5807&2006-03-24&127.0(126.9)\\  
Abell 2063&0.0354&PGC 054913&10.79 &11.67$\pm$0.02&&6263&2005-03-29&16.8(16.8)\\  
Abell 2199&0.0302&NGC 6166  &11.38 &11.90$\pm$0.01&& 497&2000-05-13&21.5(19.2)\\  
Abell 2589&0.0416&NGC 7647  &10.95 &11.85$\pm$0.02&&7190&2006-06-11&53.8(53.4)\\  
Abell 3112&0.0750&PGC 012264&11.35 &12.15$\pm$0.03&&2516&2001-09-15&17.5(16.5)\\  
Abell 3558&0.0480&PGC 047202&11.39 &12.17$\pm$0.02&&1646&2001-04-14&14.4(14.2)\\
Abell 3571&0.0397&PGC 048896&11.55 &12.08$\pm$0.01&&4203&2003-07-31&34.0(31.6)\\  
Abell 4038&0.0283&IC 5358   &10.95 &11.68$\pm$0.02&&4992&2004-06-28&33.5(33.5)\\  
Abell 4059&0.0460&PGC 073000&11.32 &12.00$\pm$0.02&&5785&2005-01-26&92.4(92.1)\\  
MS 0735.6+7421&0.2160&4C +74.13&11.11 &11.85$\pm$0.07&&4197&2003-11-30&45.9(44.9)\\  
\hline\hline
\end{tabular}
\end{center}
\tablecomments{0.9\textwidth}{
\scriptsize
$^{\it a}$$B$-band luminosities of the CDGs shown as log($L_{B}/{\rm L_{B,\odot}}$), 
which are calculated using the data drown from 
http://leda.univ-lyon1.fr (Paturel et al. 2003). 
$^{\it b}$2MASS $K$-band luminosities of the CDGs shown as log($L_{K}/{\rm L_{K,\odot}}$), 
which are calculated using the data drown from http://www.ipac.caltech.edu/2mass.
}
\end{table}

\subsection{Observation and data reduction}
All the {\it Chandra} observations (Table 1) were performed with 
the ACIS instrument operated at a focal plane temperature of $-120$ $^{\circ}$C, 
and all the selected galaxy groups and clusters were positioned 
close to the nominal aim point on CCD 7 (ACIS-S) or on CCD 3 (ACIS-I). 
We employed the {\it Chandra} data analysis package CIAO v4.1 and the 
calibration CALDB v4.1.2 to process the datasets in the 
standard way, by starting with the Level-1 event files. 
We kept events with {\it ASCA} grades 0, 2, 3, 4, and 6, removed 
all the bad pixels and bad columns, and executed corrections for both
the charge transfer inefficiency (CTI) and time dependent gain.
We selected regions located as far away from the galaxy groups and 
clusters as possible to extract the $0.3-10.0$ keV and $2.5-7.0$ keV 
lightcurves for front-illuminated CCDs (CCD $0-3$) and back-illuminated 
CCDs  (CCD 5 \& 7), respectively, and then examined if there exist occasional 
background flares, the intervals contaminated by which were excluded 
in our study. In the spectral analysis that follows, we mask all the 
X-ray point sources detected in $0.3-8.0$ keV with the CIAO tool 
{\it celldetect}, and apply the spectra extracted from {\it Chandra} 
blank-sky fields as the background. When source free regions are available 
in the observations, we also have attempted to use the spectra extracted 
from these regions as the background; we find that the results obtained 
with either background sets are consistent with each other within 
the 90\% error limits.  

\section{SPECTRAL ANALYSIS AND RESULTS}
\subsection{Model fittings}
In order to measure the spatial distribution of specific 
gas entropy, which is defined as $K=kTn_{e}^{-2/3}$, where 
$T$ and $n_{e}$ are gas temperature and electron density, respectively,
we divide each galaxy group or cluster into concentric annuli, and 
study the spectra extracted therein with XSPEC v12.4.0 by applying 
a model that consists of an APEC component to represent the gas 
emission and a power-law component to represent the emission
from unresolved point sources, both subject to a common absorption 
due to the neutral hydrogen. We set the metal abundance free to vary 
and fix the absorption at the Galactic value (Dickey \& Lockman 1990),
except that in 2A 0335+096 and Abell 478 the absorption is left free, 
because in these two galaxy clusters significant absorption excesses were 
reported in previous works (Mazzotta et al. 2003; Sanderson et al. 2005).
We adopt the {\it projct} model embedded in the XSPEC package to 
correct the projection effect. By applying the F-test, we find that 
in the central annuli of the NGC 383, NGC 741, NGC 1407, NGC 4261, 
and NGC 5129 groups the power-law component is required at $90\%$ 
confidence level, while in outer regions of these groups the power-law 
component is negligible. In other sample systems, the contribution of 
the power-law component is negligibly small. 

\begin{figure}[h!!!]
\begin{center}
\includegraphics[width=14cm,angle=0]{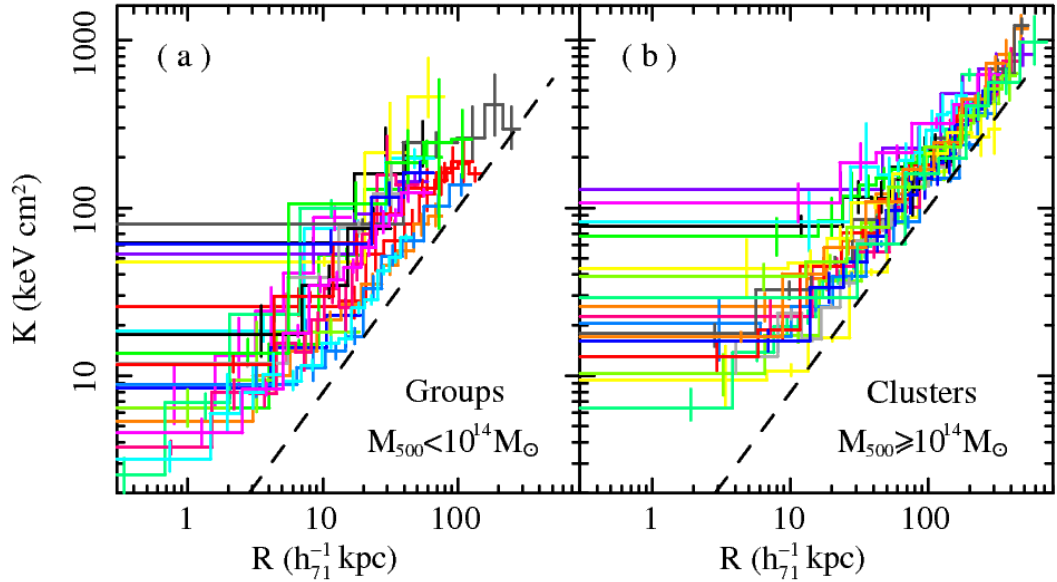}
\end{center}
\caption{
Deprojected azimuthally-averaged radial distributions of gas entropy $K(R)$ 
of the galaxy groups ($M_{500}<10^{14}$ M$_{\odot}$; a)
and galaxy clusters ($M_{500}\geq 10^{14}$ M$_{\odot}$; b)
in the sample, along with the theoretical prediction $K\propto R^{1.1}$ 
with an arbitrary normalization (dashed lines),
which is ascribed to shock heating that occurred during spherical collapses
(Voit $\&$ Donahue 2005).
}
\end{figure}

Using the best-fit deprojected spectral parameters, we calculate 
the 3-dimensional azimuthally-averaged radial entropy distribution 
$K(R)$ for all the sample systems and plot them in Figure 1. 
Following Donahue et al. (2006), we fit the obtained radial entropy 
profiles with a three-parameter expression 
\begin{equation}
K(R)=\Delta K_{0}+K_{100}(\frac{R}{100h_{71}^{-1}~{\rm kpc}})^{\alpha},
\end{equation}
where $\Delta K_{0}$ represents the central gas entropy excess
beyond the best-fit power-law model for larger radii,
$K_{100}$ is the normalization of the power-law component at 100$h_{71}^{-1}$ kpc, 
$\alpha$ is the power-law index, and $R$ is the 3-dimensional radius. 
All of the best-fit parameters are listed in Table 2.
In all systems the gas entropy profile is well fitted with a 
power-law model in outer regions. In nine systems the power-law model 
can be extrapolate towards the center, resulting in a good acceptable fit. 
In other 31 systems, however, there exists a significant central 
($R\lesssim 10h_{71}^{-1}$ kpc) entropy excess beyond the best-fit 
power-law model for outer regions. The obtained central entropy 
excesses vary in the range of $0-100$ keV cm$^{2}$. 
This entropy excess is usually mentioned as the central gas entropy 
plateau, and is suspected to be produced by heating sources like AGNs 
(e.g., Voit \& Donahue 2005; McNamara \& Nulsen 2007).  
Note that, in 2A 0335+096, Abell 85, Abell 496, Abell 3558, Abell  
3571, and Abell 4059 clusters,
the evidence of minor merger has been reported in literature 
(Werner et al. 2006; Durret et al. 2005; Tanaka et al. 2006; 
Rossetti et al. 2007; Hudaverdi et al. 2005; Choi et al. 2004),
which possibly result in mixing of high and low entropy gas
(e.g., Ricker \& Sarazin 2001; McCarthy et al. 2007).
For these clusters, the average slope of the entropy profiles  
($1.13\pm0.17$) shows no significant systematic bias from those of the sample 
($1.07\pm0.28$), which indicate that minor merger likely does not  
appreciably affect the gas entropy distribution of the sample 
groups and clusters (Ghizzardi et al. 2010).

\subsection{Correlation between central gas entropy excess and CDG's K-band luminosity}

In order to investigate the possible relation between the central 
gas entropy excess and the AGN heating, we apply the near-infrared 
$K$-band luminosity $L_{K}$ of the sample CDG by adopting the apparent 
$K$-band magnitudes of the CDGs from the Two Micron All Sky Survey 
(2MASS) archive\footnote[1]{See http://www.ipac.caltech.edu/2mass.}
and converting them into $K$-band luminosities (Table 1).
In the calculation, we have corrected Galactic extinction 
(Schlegel et al. 1998), and applied the correction for redshift 
(i.e., $k$-correction) as $k(z)=-6$log$(1+z)$ (Kochanek et al. 2001). 

In Figure 2, we show the central gas entropy excess $\Delta K_{0}$
versus the $K$-band luminosity $L_{K}$ for the 40 CDGs in our sample. 
It can be seen that $\Delta K_{0}$ apparently increases from 0 to 100 
keV cm$^2$ as $L_{K}$ increases from $1\times10^{11}$ to 
$3\times10^{12}$ ${\rm L_{K,\odot}}$, and $\Delta K_{0}$ shows a roughly uniform 
scatter of $\simeq0.5$ in logarithmic scale. We find that the correlation 
coefficient for $\Delta K_{0}$ and $L_{K}$ is 0.45, which indicates 
that the two parameters has a correlation at $99\%$ confidence level. 
We also have calculated the Kolmogorov-Smirnov (K-S) statistic of 
$\Delta K_{0}$ and $L_{K}$ against a proper constant model, 
and find that the correlation has a probability of more than 90\% 
if the null hypothesis is true. Using the bisector of ordinary 
least-squares regression (Isobe et al. 1990), which is suitable 
to fit data with large scatters, we fit the log $\Delta K_{0}-$log $L_{K}$
relation with a liner model of 
\begin{equation}
{\rm log} (\Delta K_{0})=A+B~[{\rm log}({L_{K}}/{\rm L_{K,\odot}})-10.9],
\end{equation}
and obtain $A=-0.8\pm0.3$ and $B=1.6\pm0.4$ (Fig. 2), where the 
errors are determined by performing the Jackknife simulation 
that has been repeated 100 times (Feigelson \& Babu 1992).

\begin{figure}[h!!!]
\begin{center}
\includegraphics[width=13cm,angle=0]{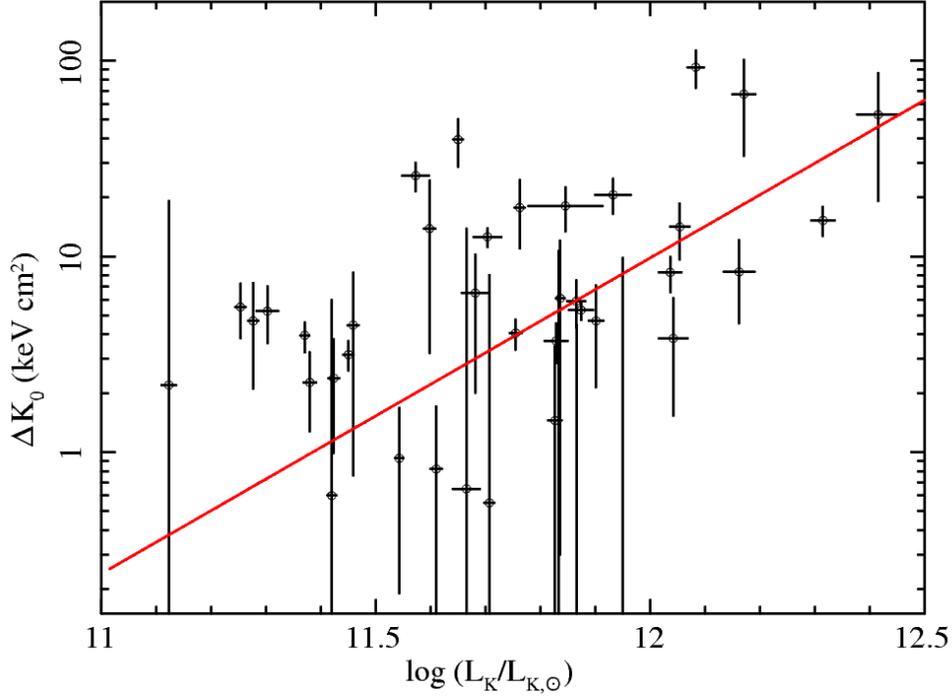}
\end{center}
\caption{
Central gas entropy excess $\Delta K_{0}$ vs $K$-band luminosity $L_{K}$ 
for the 40 CDGs of our sample (see also Table 1 and 2). 
The solid line shows the best-fit model  
${\rm log} (\Delta K_{0})=-0.8\pm0.3+(1.6\pm0.4)[{\rm log}({L_{K}}/{\rm L_{K,\odot}})-10.9]$ 
(Eq. 2 and \S3.2), which is determined by the bisector of ordinary 
least-squares regression (Isobe et al. 1990). 
}
\end{figure}

\begin{landscape}
\begin{table}[h!!!]
\caption{Deprojected spectral analysis using an absorbed APEC+POWERLAW model}
\begin{center}
\tiny
\begin{tabular}{lccccccccccc}
\hline\hline
Group/cluster&$N_{H}$$^{~\it a}$&$\Delta K_{0}$$^{~\it b}$&$\alpha$$^{~\it  
c}$&$K_{100}$$^{~\it d}$&
$T_{X}$$^{~\it e}$&$M_{500}$$^{~\it f}$&$R_{\rm cool}$$^{~\it g}$&
log ($L_{X}$)$^{~\it h}$&
$E_{\rm feedback}^{AGN}$$^{~\it i}$&$E_{\rm feedback}^{SN}$$^{~\it  j}$&$E_{\rm cool}$$^{~\it k}$\\
&($10^{20}$ $cm^{-2}$)&(keV cm$^{2}$)&&(keV cm$^{2}$)&
(keV)&$10^{14}$ M$_{\odot}$&($h_{71}^{-1}$ kpc)&(ergs s$^{-1}$)&
(keV/gas particle)&(keV/gas particle)&(keV/gas particle)\\
\hline
NGC 0383 &5.39&$0.55^{+7.47}_{-0.55}$&$0.75\pm0.05$&$389\pm47$&
$2.14^{+1.49}_{-0.60}$&0.72&11&$41.06\pm0.05$&
2.26&0.0047&0.002$\pm0.0001$\\
NGC 0507 &5.24&$17.8\pm6.8$&$1.40^{+0.31}_{-0.22}$&$179\pm37$&
$1.55\pm0.06$&0.42&52&$42.40^{+0.09}_{-0.06}$&
4.76&0.0123&0.069$\pm0.0002$\\
NGC 0533 &3.10&$4.06^{+0.69}_{-0.72}$&$1.43\pm0.04$&$588\pm50$&
$1.72^{+0.65}_{-0.36}$&0.50&25&$42.03^{+0.09}_{-0.12}$&
3.85&0.0115&0.024$\pm0.0001$\\
NGC 0720 &1.54&$5.52^{+1.74}_{-1.69}$&$0.91^{+0.19}_{-0.11}$&$100\pm44$&
$0.44^{+0.16}_{-0.21}$&0.05&40&$40.77^{+0.22}_{-0.21}$&
13.29&0.0385&0.017$\pm0.0001$\\
NGC 0741 &4.43&$6.11^{+5.93}_{-5.81}$&$1.07^{+0.10}_{-0.08}$&$420\pm70$&
$1.80^{+0.23}_{-0.21}$&0.54&24&$41.62^{+0.06}_{-0.09}$&
4.37&0.0139&0.009$\pm0.0001$\\
NGC 1407 &5.42&$4.46^{+3.80}_{-3.70}$&$1.36^{+0.12}_{-0.09}$&$1440\pm340$&
$1.27^{+0.34}_{-0.18}$&0.30&16&$40.81^{+0.15}_{-0.32}$&
3.14&0.0102&0.003$\pm0.0001$\\
NGC 2563 &4.23&$0.60^{+5.40}_{-0.60}$&$1.30\pm0.05$&$900\pm100$&
$1.69^{+0.18}_{-0.11}$&0.49&17&$41.17\pm0.07$&
1.66&0.0039&0.003$\pm0.0001$\\
NGC 3402 &4.43&$2.39^{+1.38}_{-1.40}$&$0.88\pm0.06$&$122\pm11$&
$0.93\pm0.04$&0.18&51&$42.49\pm0.03$&
5.12&0.0179&0.218$\pm0.0002$\\
NGC 4125 &1.84&$4.68^{+2.63}_{-2.57}$&$0.84^{+0.21}_{-0.14}$&$83\pm33$&
$0.40^{+0.20}_{-0.18}$&0.05&33&$40.55^{+0.23}_{-0.36}$&
16.88&0.0750&0.012$\pm0.0001$\\
NGC 4261 &1.55&$2.27^{+0.97}_{-0.99}$&$1.51\pm0.06$&$2480\pm360$&
$1.43^{+0.54}_{-0.21}$&0.37&12&$40.78^{+0.07}_{-0.09}$&
2.04&0.0072&0.002$\pm0.0001$\\
NGC 4325 &2.24&$5.25^{+1.77}_{-1.64}$&$1.30\pm0.09$&$141\pm15$&
$1.07^{+0.03}_{-0.02}$&0.23&58&$42.83^{+0.03}_{-0.04}$&
2.89&0.0109&0.370$\pm0.0003$\\
NGC 5044 &4.93&$3.95^{+0.65}_{-0.71}$&$1.38\pm0.03$&$208\pm8 $&
$1.30\pm0.02$&0.32&54&$42.60\pm0.01$&
2.38&0.0117&0.150$\pm0.0001$\\
NGC 5129 &1.76&$39.5\pm10.8$&$1.18\pm^{+0.24}_{-0.18}$&$247\pm55$&
$0.98^{+0.08}_{-0.05}$&0.20&16&$41.56^{+0.06}_{-0.07}$&
8.41&0.0329&0.023$\pm0.0001$\\
NGC 5846 &4.26&$3.15^{+0.55}_{-0.54}$&$1.33\pm0.07$&$431\pm63$&
$1.01\pm0.04$&0.21&22&$41.50^{+0.17}_{-0.29}$&
4.69&0.0131&0.019$\pm0.0001$\\
NGC 6269 &4.77&$0.01^{+9.81}_{-0.01}$&$0.63\pm0.08$&$226\pm26$&
$2.53^{+0.28}_{-0.36}$&0.94&31&$42.19\pm0.05$&
3.13&0.0113&0.017$\pm0.0001$\\
NGC 7619 &5.00&$0.93^{+0.75}_{-0.74}$&$1.02\pm0.02$&$424\pm21$&
$1.03\pm0.03$&0.21&16&$41.28^{+0.10}_{-0.15}$&
5.76&0.0245&0.011$\pm0.0001$\\
HCG 42   &4.81&$13.9^{+10.6}_{-10.7}$&$1.47^{+0.42}_{-0.23}$&$518\pm270$&
$0.93^{+0.10}_{-0.12}$&0.18&21&$41.48^{+0.10}_{-0.09}$&
8.06&0.0236&0.021$\pm0.0001$\\
MKW 3S   &3.03&$25.9^{+4.2}_{-4.4}$&$1.10\pm0.09$&$149\pm9 $&
$3.96\pm0.10$&1.95&106&$44.02\pm0.01$&
0.51&0.0021&0.508$\pm0.0001$\\
MKW 4    &1.89&$1.45^{+2.02}_{-1.45}$&$0.81\pm0.04$&$223\pm18$&
$2.00^{+0.07}_{-0.05}$&0.64&47&$42.76\pm0.02$&
3.50&0.0100&0.099$\pm0.0001$\\
MKW 4S   &1.69&$0.01^{+4.45}_{-0.01}$&$0.99\pm0.11$&$492\pm137$&
$1.81^{+0.75}_{-0.48}$&0.54&24&$42.08^{+0.03}_{-0.04}$&
4.69&0.0139&0.025$\pm0.0001$\\
UGC 12064&11.8&$2.20^{+17.0}_{-2.20}$&$0.66^{+0.16}_{-0.13}$&$244\pm46$&
$1.65^{+0.56}_{-0.28}$&0.47&27&$41.62^{+0.06}_{-0.07}$&
0.80&0.0032&0.010$\pm0.0001$\\
2A0335+096&20.3(1.78)&$3.70\pm0.85$&$1.43\pm0.04$&$141\pm5 $&
$4.22^{+0.16}_{-0.13}$&2.18&131&$44.40^{+0.03}_{-0.05}$&
0.88&0.0029&1.072$\pm0.0012$\\
Abell 0085&3.42&$14.1\pm4.5$&$1.09\pm0.08$&$174\pm12$&
$6.32\pm0.17$&4.21&120&$44.44^{+0.03}_{-0.06}$&
0.75&0.0014&0.556$\pm0.0008$\\
Abell 0262&5.38&$0.82^{+0.89}_{-0.82}$&$1.09\pm0.03$&$304\pm19$&
$2.39\pm0.05$&0.86&47&$42.87^{+0.07}_{-0.05}$&
1.43&0.0021&0.091$\pm0.0001$\\
Abell 0478&28.5(1.51)&$3.81^{+2.32}_{-2.27}$&$1.06\pm0.05$&$150\pm7 $&
$7.50^{+0.38}_{-0.36}$&5.49&182&$45.17^{+0.04}_{-0.08}$&
0.53&0.0013&2.218$\pm0.0039$\\
Abell 0496&4.57&$5.89^{+1.66}_{-1.56}$&$1.15\pm0.05$&$178\pm8 $&
$5.29\pm0.09$&3.17&109&$44.12\pm0.04$&
0.63&0.0018&0.369$\pm0.0003$\\
Abell 0780&4.92&$12.6^{+1.3}_{-1.4}$&$1.09\pm0.04$&$107\pm4 $&
$3.47^{+0.05}_{-0.06}$&1.57&155&$44.47^{+0.02}_{-0.03}$&
0.92&0.0062&1.810$\pm0.0012$\\
Abell 1651&1.81&$53.0^{+33.4}_{-33.8}$&$0.80\pm0.14$&$234\pm38$&
$7.45^{+0.83}_{-0.77}$&5.45&96&$44.24\pm0.02$&
1.42&0.0006&0.258$\pm0.0001$\\
Abell 1795&1.19&$20.7\pm4.2$&$1.32\pm0.10$&$127\pm10$&
$6.30\pm0.20$&4.17&148&$44.74^{+0.03}_{-0.06}$&
0.55&0.0009&1.104$\pm0.0015$\\
Abell 2029&3.05&$15.3\pm2.6$&$1.06\pm0.04$&$169\pm6 $&
$8.10\pm0.15$&6.28&160&$45.02^{+0.03}_{-0.05}$&
0.93&0.0011&1.338$\pm0.0015$\\
Abell 2052&2.73&$5.32^{+0.48}_{-0.58}$&$1.48\pm0.02$&$210\pm8 $&
$3.20\pm0.04$&1.38&109&$43.95^{+0.04}_{-0.06}$&
1.65&0.0022&0.634$\pm0.0009$\\
Abell 2063&2.98&$0.65^{+13.2}_{-0.65}$&$0.37\pm0.12$&$170\pm16$&
$3.70\pm0.15$&1.75&90&$43.64\pm0.01$&
0.75&0.0014&0.239$\pm0.0001$\\
Abell 2199&0.86&$4.68^{+2.42}_{-2.53}$&$0.97\pm0.06$&$181\pm9 $&
$4.80\pm0.14$&2.70&120&$44.22^{+0.07}_{-0.14}$&
0.83&0.0034&0.557$\pm0.0018$\\
Abell 2589&4.14&$0.01^{+10.7}_{-0.01}$&$0.45\pm0.10$&$208\pm13$&
$3.64\pm0.10$&1.70&75&$43.52\pm0.01$&
1.21&0.0021&0.185$\pm0.0001$\\
Abell 3112&2.61&$8.33^{+3.79}_{-3.77}$&$1.12\pm0.13$&$155\pm20$&
$5.56^{+0.30}_{-0.25}$&3.37&138&$44.58\pm0.01$&
1.23&0.0024&0.984$\pm0.0002$\\
Abell 3558&3.89&$67.4^{+33.6}_{-34.7}$&$1.21\pm0.46$&$199\pm41$&
$7.69\pm0.41$&5.83&86&$43.92\pm0.01$&
0.70&0.0014&0.114$\pm0.0003$\\
Abell 3571&3.71&$92.6^{+20.1}_{-20.0}$&$0.94\pm0.12$&$193\pm17$&
$7.17\pm0.23$&5.21&101&$44.26^{+0.01}_{-0.04}$&
0.63&0.0024&0.290$\pm0.0003$\\
Abell 4038&1.56&$6.50^{+3.71}_{-4.49}$&$0.73\pm0.10$&$186\pm14$&
$3.24\pm0.11$&1.41&78&$43.41\pm0.01$&
0.98&0.0028&0.180$\pm0.0001$\\
Abell 4059&1.10&$8.28^{+1.64}_{-1.72}$&$0.96\pm0.05$&$182\pm6 $&
$4.36^{+0.09}_{-0.08}$&2.29&101&$43.94^{+0.08}_{-0.22}$&
1.30&0.0035&0.349$\pm0.0018$\\
MS 0735.6+7421&3.49&$18.1^{+4.5}_{-4.7}$&$1.26\pm0.14$&$112\pm17$&
$5.32^{+0.41}_{-0.32}$&2.94&125&$44.56\pm0.01$&
0.65&0.0014&1.079$\pm0.0003$\\
\hline\hline
\end{tabular}
\end{center}
\tablecomments{1.3\textwidth}{
\scriptsize
$^{\it a}$Absorptions are fixed to Galactic values (Dickey \& Lockman 1990),
except for those of 2A 0335+096 and Abell 478,
which are left free and shown in brackets (see \S3.1 for details).
$^{\it b-d}$Central gas entropy excesses, power-law indices,
and normalizations of Eq. 1 (\S3.1).
$^{\it e}$Mean gas temperatures are measured from the spectra extracted
within $0.1-0.2r_{500}$. For each system,
the mean gas temperature and $r_{500}$ follow the relation
$r_{500}=391\times T_{X}^{0.63}/E(z)$ kpc (Willis et al. 2005).
$^{\it f}$Total gravitating masses within $r_{500}$ (\S4).
$^{\it g}$Cooling radii, at which the cooling time equals the universe's age
at the system's redshift.
$^{\it h}$$0.3-12.0$ keV luminosities are measured within the cooling radius
in logarithmic scale.
$^{\it i}$Cumulative AGN feedback to heating the IGM (\S4 and Fig. 3), 
which is assumed $E^{\rm AGN}_{\rm feedback}\simeq\eta M_{\rm BH}c^{2}$ and $\eta=0.02$. 
$^{\it j}$Supernova feedback to heating the IGM (\S4 and Fig. 3), 
which includes both type Ia and II supernova contributions.
$^{\it k}$X-ray Radiative loss since $z=2$ (\S4 and Fig. 3), 
which is calculated from the $0.3-12$ keV luminosity within 
the cooling radius ($R_{\rm cool}$).
}
\end{table}
\end{landscape}

\section{DISCUSSION}
By analyzing the deprojected gas entropy profiles, we find that there 
exists a significant central gas entropy excess in 78\% sample groups 
and clusters, which can be ascribed to the non-gravitational heating processes. 
The average central entropy excess ranges from 5.7 keV cm$^{2}$ to 19.3 keV cm$^{2}$ 
from groups to clusters, which corresponds to a gas energy excess of 
$\simeq0.1-0.2$ and $\simeq0.3-0.5$ keV per gas particle, respectively,
when either isodensity or isobaric heating process is assumed 
(e.g., Lloyd-Davies et al. 2000). Because the observed central 
entropy excess is correlated to the CDG's $K$-band luminosity, 
two of the most possible heating sources are AGN activity and 
supernova explosions. Here we compare their contributions 
and determine which one dominates the gas heating progress and 
is thus responsible for the observed central entropy excess.
In the calculation that follows, we adopt gravitating mass $M_{500}$ 
and gas mass fraction $f_{\rm gas,500}$ that are determined at 
$r_{500}$ ($r_{500}$ is the radius within which the over-density is 
500 with respect to the universe's critical density at each system's 
redshift) following 
$M_{500}=E(z)^{-1} 10^{14.10}(kT_{X}/3.0~{\rm keV})^{1.65\pm0.04}~{\rm M_{\odot}}$
(Sun et al. 2009) and 
$f_{\rm gas}=0.0708~(kT_{X}/1.0~{\rm keV})^{0.22}$
(Sun et al. 2009; Pratt et al. 2009), respectively, where 
$E(z)=\sqrt{(1+z)^{3}\Omega_{\rm M}+\Omega_{\Lambda}}$, 
and $T_{X}$ is the mean gas temperature measured in $0.1-0.2r_{500}$ (Table 2).

\begin{figure}[h!!!]
\begin{center}
\includegraphics[width=13cm,angle=0]{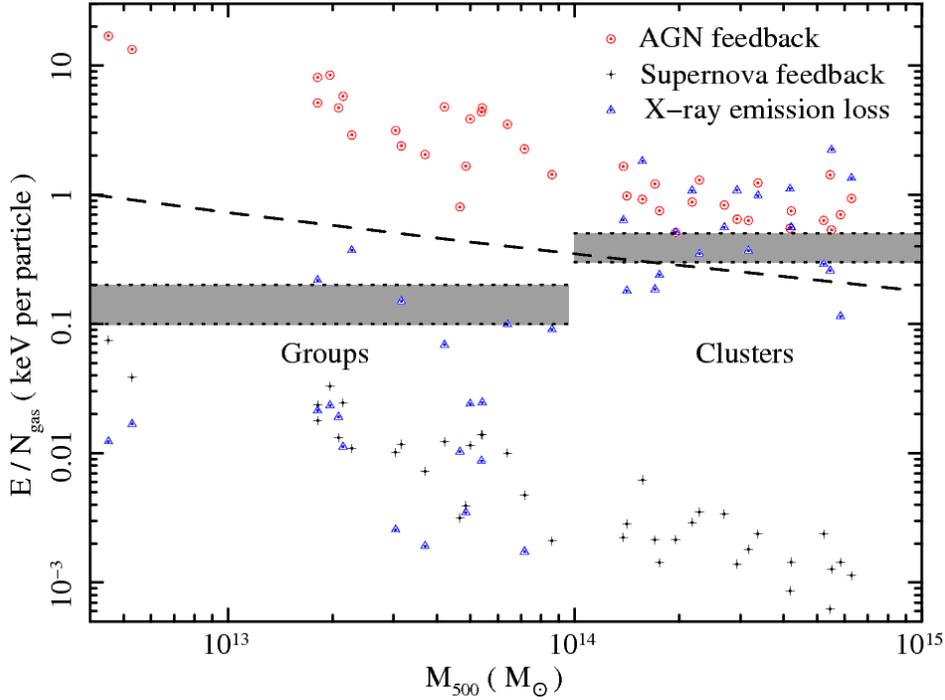}
\end{center}
\caption{ 
Estimated energy feedbacks to the IGM by AGNs (circle) and supernova
explosions (cross), comparing to the average gas energy excess of 
$\simeq0.1-0.2$ and $\simeq0.3-0.5$ keV per particle for galaxy 
groups and clusters (grey belts), respectively, the X-ray radiative 
loss since $z=2$ (triangle), and the energy required to 
deviate scaling relations from self-similar predictions 
(dashed line; Bode et al. 2009; see \S4 for details). 
}
\end{figure}

\noindent{\bf AGN activity}

It has been estimated that one powerful AGN outburst can 
produce a stable entropy excess of $\simeq 10-30$ keV cm$^2$ 
for $\sim10^{8}$ yr (Voit \& Donahue 2005).
In order to obtain the cumulative contribution of AGN feedback 
that can be approximated by $E^{\rm AGN}_{\rm feedback}\simeq\eta M_{\rm BH}c^{2}$
(e.g., Croton et al. 2006; Short \& Thomas 2009), 
we calculate the masses of the central black holes hosted in the sample CDGs 
by applying the relation between the near-infrared $K$-band luminosity $L_{K}$ 
of the host galaxy and the black hole mass $M_{\rm BH}$, i.e.,
${\rm log}(M_{\rm BH}/{\rm M_{\odot}})=8.21\pm0.07+(1.13\pm0.12)
\times[{\rm log}(L_{K}/{\rm L_{K,\odot}})-10.9]$, which is presented in 
Marconi \& Hunt 2003 and shows similar scatter ranges to the 
more usually used relation between the stellar velocity dispersion 
of the galaxy's bulge and $M_{\rm BH}$ 
(Marconi \& Hunt 2003; Batcheldor et al. 2007).
We adpot a conversion efficiency of $\eta=0.02$, because in the study of 
X-ray cavities in nine elliptical galaxies 
Allen et al. (2006) found that about $2\%$ of the accreting energy
is converted into the thermal energy of the surrounding gas. 
By assuming that each gas particle within $r_{500}$ 
of galaxy groups and clusters has been heated uniformly, 
we find that the cumulative contribution of AGN feedback 
varies between $\simeq0.5-17.0$ keV per particle for our sample, 
which decreases as the gravitating mass $M_{500}$ increases (Table 2 and Fig. 3),
and is apparently higher than the average gas energy excess 
($\simeq0.1-0.5$ keV per particle).

\noindent{\bf Supernova explosions} 

Using CDGs' $B-$band luminosities ($L_{B}$; Table 1) and observationally 
constrained explosion rates of type Ia supernova 
(SN Ia; Dahlen et al. 2004; Cappellaro et al. 2005)
and adopting that the supernova explosions heat the surrounding gas 
with $\sim 10^{50}$ ergs per event (e.g., Thornton et al. 1998), 
we find that the contribution of SN Ia feedback to the IGM since $z=3$ 
(i.e., about $10-11$ Gyr ago) is of the order of 
$\simeq10^{-4}-10^{-2}$ keV per particle.
On the other hand, we note that, although no type II supernova (SN II)
explosions is detected in nearby elliptical galaxies
(Cappellaro et al. 1999), the feedback energy of SN II exploded 
in high redshift and in galaxy's starburst epoch 
possibly has been deposited and contributed to
the gas heating (e.g., Bryan 2000; Wu et al. 2002).
Both field supernova observations in $z=0.1-0.9$ 
(Dahlen et al. 2004; Cappellaro et al. 2005)
and galaxy star formation theories based on initial stellar mass functions 
(e.g., Kravtsov \& Yepes 2000) indicate that 
the explosion ratio of SN II to SN Ia is $\simeq2-3$.
This value agrees with what were obtained in the studies on IGM metallicity 
(e.g., Wang et al. 2005; De Grandi \& Molendi 2009), which suggested that the 
explosion ratio of SN II to SN Ia is $\simeq1-6$, 
the larger scatter of which is probably due to the clusters' different
merger history. By assuming that the feedback energy per SN II event is 
the same as that of SN Ia event (e.g., Woosley \& Weaver 1986), and 
that the cumulative explosion ratio of SN II to SN Ia is 3, 
we estimate that the total contribution of supernova feedback to the IGM 
is $\simeq0.0006-0.08$ keV per particle, which also decreases 
as the gravitating mass $M_{500}$ becomes larger (Table 2 and Fig. 3). 
The estimated supernova contribution is apparently lower 
than the average gas energy excess ($\simeq0.1-0.5$ keV per particle), 
and mostly is about 1-2 orders of magnitude lower than 
the average gas energy excess. For all the sample groups and clusters, 
it is about 2-3 orders of magnitude lower than the AGN feedback.

In most case the estimated AGN contribution can compensate 
the radiative loss since $z=2$ (i.e., about $9-10$ Gyr ago) 
in galaxy clusters ($\simeq0.1-2.2$ keV per particle; Table 2 and Fig. 3), 
which is calculated from the $0.3-12$ keV luminosity within 
the cooling radius (Table 2). For galaxy groups ($<10^{14}~{\rm M_{\odot}}$), 
the AGN feedback energy is about $1-2$ orders of magnitude higher than the 
X-ray radiative loss ($\simeq0.002-0.4$ keV per particle; Fig. 3). 
Moreover, this surplus energy fed by AGNs in galaxy groups is expected
to re-distribute the IGM gas, especially in the central regions, 
and thus break the self-similarity between galaxy groups and clusters.  
Assuming a uniform baryon 
(mainly includes gas and stellar components) mass fraction 
for galaxy groups and clusters, the energy required 
to deviate the scaling relations from self-similar predictions 
is $\simeq0.2-1.0$ keV per particle (Bode et al. 2009), 
which can be supplied by AGN not by supernova explosions.

\section{SUMMARY}
In 31 galaxy groups and clusters we find that there exists 
a significant central gas entropy excess,  
which scales with the $K$-band luminosity of the CDG via 
$\Delta K_{0}\propto L_{K}^{1.6\pm0.4}$. 
By comparing between the contributions of AGN activity and 
supernova explosions, we conclude that AGNs are 
responsible for the central entropy excesses.
  
\section*{Acknowledgments}
We thank the {\it Chandra} team for making data available via the High
Energy Astrophysics Science Archive Research Center (HEASARC). 
This work was supported by the National Science Foundation of China 
(Grant No. 10673008, 10878001 and 10973010), the Ministry of 
Science and Technology of China (Grant No. 2009CB824900/2009CB24904), 
and the Ministry of Education of China (the NCET Program).

\label{lastpage}

\end{document}